\documentclass[english,usenatbib]{mn2e}
\usepackage[]{fontenc}
\usepackage[latin1]{inputenc}
\usepackage{amsmath}
\usepackage{graphicx}
\usepackage{amssymb}
\usepackage[authoryear]{natbib}

\newcommand{\rcore}{\mbox{${R_{\rm core}}$}}

\newcommand{\trlx}{\mbox{${t_{\rm rh}}$}}
\newcommand{\tcross}{\mbox{${t_{\rm cross}}$}}
\newcommand{\tnb}{\mbox{${t_{\rm Nb}}$}}

\newcommand{\msun}{\mbox{${\rm M}_\odot$}}
\newcommand{\rsun}{\mbox{${\rm R}_\odot$}}
\newcommand{\tdecay}{\mbox{$t_{\rm decay}$\,}}
\newcommand{\tcoll}{\mbox{$t_{\rm coll}$\,}}

\newcommand{\tcc}{\mbox{$t_{\rm cc}$\,}}

\newcommand{\rvir}{\mbox{$R_{\rm vir}$\,}}

\newcommand{\mf}{\mbox{${\cal N}$}}
\newcommand{\eq}[1]{\mbox{Eq. #1}}
\newcommand{\fig}[1]{\mbox{Fig. #1}}
\newcommand{\tbl}[1]{\mbox{Tab. #1}}
\newcommand{\sect}[1]{\mbox{Sect. #1}}
\newcommand{\append}[1]{\mbox{Appendix #1}}

\newcommand{\mean}[1]{\mbox{$\langle #1 \rangle$}}

\usepackage{babel}

\title[Dynamics of the first collision]
      {On the onset of runaway stellar collisions in dense star clusters
        I. Dynamics of the first collision.}

\author[E. Gaburov, A. Gualandris and S. Portegies Zwart]
       {E. Gaburov,$^{1,2}$\thanks{E-mail: egaburov@science.uva.nl
           (EG); alessiag@astro.rit.edu (AG); spz@science.uva.nl
           (SPZ))} A. Gualandris, $^{1,2,3}$\footnotemark[1] and
         S. Portegies Zwart$^{1,2}$\footnotemark[1]\\
       $^1$ Astronomical Institute 'Anton Pannekoek' University of Amsterdam, the Netherlands \\ 
       $^2$ Section Computational Science, University of Amsterdam, the Netherlands \\ 
       $^3$Center for Computational Relativity and Gravitation, \\
         $\qquad$ Rochester Institute of Technology, 85 Lomb Memorial Drive, Rochester, NY 14623, USA\\
       }

\begin{document}

\maketitle
\date{Accepted XXX. 
      Received XXX
     }
\pagerange{\pageref{firstpage}--\pageref{lastpage}} 
\pubyear{2007}
\label{firstpage}

\begin{abstract}
  We study the circumstances under which first collisions occur in young
  and dense star clusters. The initial conditions for our direct
  $N$-body simulations are chosen such that the clusters experience core
  collapse within a few million years, before the most massive stars
  have left the main-sequence. It turns out that the first collision is
  typically driven by the most massive stars in the cluster. Upon
  arrival in the cluster core, by dynamical friction, massive stars tend
  to form binaries. The enhanced cross section of the binary compared to
  a single star causes other stars to engage the binary.  A collision
  between one of the binary components and the incoming third star is
  then mediated by the encounters between the binary and other cluster
  members.  Due to the geometry of the binary-single star engagement the
  relative velocity at the moment of impact is substantially different
  than in a two-body encounter. This may have profound consequences for
  the further evolution of the collision product.
\end{abstract}

\begin{keywords}
  methods: N-body simulations -- young star clusters
\end{keywords}

\section{Introduction}\label{sect:Introduction}

In recent years, it became clear that stellar collisions are common
events in young dense star clusters, and that such events are natural
ways to form stellar exotica. In extreme cases, it is even possible
that a large number of stars merge to form a very massive object. This
object can potentially be a progenitor of an intermediate mass black
hole.

\cite{1999A&A...348..117P} carried out the first $N$-body simulations
of runaway stellar collisions. In these simulations, a very massive
object forms in a young dense star cluster in just a few million
years. It was found that the collision rate is roughly an order of
magnitude greater than one would naively expect from collision
cross-section arguments. The cause of the discrepancy is mass
segregation, which enhances the central region with massive
stars. Once in the core, these stars dominate the collision rate
because of their large masses and radii. Since a collision occurs
preferentially between two massive stars, the collision product
becomes one of the most massive objects in the central region. If
conditions are right, the product can experience multiple collisions,
each time increasing its mass. This process can lead to the formation of
a very massive object. The evolution of very massive stars and of collision
products is not yet well understood, but one may speculate that
these massive objects evolve into intermediate mass black holes.

Numerical simulations carried out by \cite{2004Natur.428..724P} show
that the onset of a runaway merger depends on both the dynamical
friction timescale and the central concentration of the star
cluster. In particular, a necessary condition for the runaway
collision to proceed is the dynamical friction timescale to be smaller
than the lifetime of massive stars, which is about a few million
years. If this condition is not satisfied, the mass loss due to
supernovae causes the cluster to expand, and this prevents the cluster
from developing the high densities required for subsequent collisions.

In their work, \cite{2004ApJ...604..632G} carried out systematic
studies of mass segregation and core collapse in dense star
clusters. They found that moderately concentrated star clusters with a
realistic initial mass function can reach the runaway phase before
massive stars produce supernovae. The runaway phase was also studied
by \cite{2006MNRAS.368..141F}, who found that the mass function of
colliding stars is bimodal. The first peak lies in the lower limit of
the initial mass function, 0.1 - 1\msun, whereas the second peak is
close to the high-mass end of the initial mass function,
40-120\msun. They also found that collisions occur every few ten
thousand years, which is roughly the time required for a collision
product to reach the main-sequence.

Stellar collisions also play an important role in the formation of
stellar exotica in young star clusters, such as blue stragglers. If
cluster properties are such that the runaway phase is not possible,
stellar collisions might still occur in the cluster producing massive
bright stars. Such stars might easily be misclassified during
observations.  A possible example of these stars is provided by the
Pistol star in the Quintuplet cluster \citep{1998ApJ...506..384F},
which is thought to have an initial mass in excess of two hundred
solar masses. The lifetime of such a star is roughly three million
years. However, the cluster population is about six million years old,
significantly older than the Pistol star. It is therefore possible
that the Pistol star is in fact an ejected collision product instead
of a primordial very massive star.

In this paper, we study the dynamics of the first stellar collision in
young star clusters. The aim is to find a set of appropriate initial
conditions for subsequent hydrodynamic simulations of collision
products. In particular, we focus on the conditions under which the
first collision takes place in a cluster, like the time and place of
the first collision, the number of stars and binaries involved, the
masses of the participating stars, the orbital parameters of the
binaries, the typical impact parameter and relative velocity in the
collision. The knowledge of the dynamical properties of stellar
collisions will allow us to perform hydrodynamic simulations of
mergers. The study of the evolution of merger products determines the
observational properties of these objects, thus providing valuable
information for their identification. Once the main characteristics of
the evolution of collision remnants will be understood, we will
proceed with the study of the dynamics of repeated stellar collisions
in star clusters, with the aim to answer the question of whether a
runaway process can result in the formation of a very massive object,
which in turn may evolve into an intermediate mass black hole.

The paper is organised as follows. In \sect{\ref{sect:SetupInitCond}}
we present initial conditions for our simulations. The detailed
studies and the geometry of the first collision is presented in
\sect{\ref{sect:Circum}} and \sect{\ref{sect:CollisionGeometry}}. A
discussion of results is presented in \sect{\ref{sect:Discussion}}.

\section{Setup and initial conditions}\label{sect:SetupInitCond}

In order to study the onset of the first collision, we focus on young
star clusters with different initial virial radii (\rvir) which we
vary over more than one order of magnitude, while maintaining the
total mass of the cluster constant. The choice of keeping $\rvir$ as a
free parameter is motivated by the aim to study the dependence of the
moment of the first collision on the cluster size.
Clusters with similar initial conditions but different sizes exhibit
homologous evolution as far as non-dynamical processes, like stellar
and binary evolution, are not of significant importance. However, in
the case of stellar collisions individual stellar radii play a crucial
role. In this case, the evolution of the star cluster in principle is
not homologous; in other words, the two stars which collide in one
case will not necessary collide if the cluster size is changed. The
reason is that by scaling the cluster in size, the stellar radii
relative to the cluster size also changes. By varying $\rvir$ we will
be able to determine its influence on the collisions in star clusters.

All our calculations are performed using the {\tt kira} integrator
from the {\tt starlab} gravitational $N$-body environment
(\cite{2001MNRAS.321..199P}, {\tt http://www.manybody.org/starlab}).
Stellar evolution in the simulations is included via the {\tt SeBa}
package \cite{1996A&A...309..179P}. Binaries, though initially not
present, form dynamically in the course of the simulations and are
evolved using {\tt SeBa}. All $N$-body simulations were carried out on
the MoDeStA ({\tt http://modesta.science.uva.nl}) cluster of GRAPE-6
\citep{2003PASJ...55.1163M, 1997ApJ...480..432M} in Amsterdam.

We present the initial conditions for the different sets of
simulations in \tbl{\ref{tbl:Initial_conditions}}.
\begin{table*}
  \begin{tabular}{lllllll}
    \hline 
    Model & $\rvir$ [pc] & $N_{\rm run}$ & \rcore [pc] & $\rho_{\rm core}$ [\msun/pc$^3$]&  \tcross [kyr] & \trlx [Myr] \\ 
    \hline 
    W9R05    & 0.05    & 100 & $3.2\cdot 10^{-3}$ & $5.5\cdot 10^{9}$ & 1.5 & 0.47 \\ 
    W9R10    & 0.10    & 100 & $6.4\cdot 10^{-3}$ & $6.8\cdot 10^{8}$ & 4.3 & 1.4 \\ 
    W9R25    & 0.23    & 100 & $0.015$            & $5.6\cdot 10^{7}$ & 15 & 4.7 \\ 
    W9R50    & 0.50    &  99 & $0.032$            & $5.5\cdot 10^{6}$ & 48 & 15 \\ 
    W9R75    & 0.75    & 110 & $0.048$            & $1.6\cdot 10^{6}$ & 88 & 28
 \\ 
    \hline
  \end{tabular}
  \caption{ Parameters of the five sets of simulations.  In each case
    the total mass of the cluster is $M\simeq1.2\times 10^4\,\msun$
    and the total number of stars is 24576.  Runs differ only in the
    choice of the half-mass radius. In the first three columns we
    report the name of the set of simulations, the half-mass radius
    (in parsec) and the number of simulations performed with these
    parameters. In the subsequent columns we give the initial core
    radius (in parsec), the initial core density (in solar masses per
    cubic parsec), the half-mass crossing time (in units of 1000
    years) and the relaxation time (in Myr).}
  \label{tbl:Initial_conditions}
\end{table*}
Each simulation is carried out with $N$=24576 single stars distributed
in a \cite{1966AJ.....71...64K} model with a scaled central potential
$W_0 = 9$ \citep{1987gady.book.....B}. We did not include primordial
binaries because we aim to study large parameter space in this work.

%%
%% +++ v1 +++
Such a choice of $W_0$ is motivated by our interest in studying young
star clusters such as R136, MGG11 and Arches
\citep{1998ApJ...493..180M, 2002ApJ...581..258F, 2005ApJ...621..278M},
and young star clusters are thought to be born with high concentration
\citep{2004ApJ...608L..25Mb}. In addition, high concentration is a
necessary condition for clusters which can experience runaway stellar
mergers \citep{2004Natur.428..724P}. As we are interested in the
internal dynamics, the effect of the tidal fields is expected to be
negligible \citep{2007MNRAS.378L..29P}; therefore, our simulations are
carried out without tidal fields.
After generating stellar positions and velocities, we assign masses to
each of the stars from an initial mass
function(\cite{1993MNRAS.262..545K, 2001MNRAS.322..231K}) between
0.1\,\msun\, and 100\,\msun, which we refer to as IMF. We subsequently
scale the velocities of all stars to bring the cluster into virial
equilibrium.
%% +++ v1 +++
Such initial conditions produce star clusters with a total mass of
roughly $10^4$\msun, which approximates well the Arches cluster
\citep{1999ApJ...525..759F, 2002ApJ...581..258F}.

For each set of initial conditions with $W_0=9$, we generate about a
hundred realisations, each of which we run until the first collision
occurs. We identify a collision in our simulations when two stars pass
each other within a distance smaller than the sum of their radii;
tidal captures are ignored. The binary evolution package allows for
semi-detached and contact binaries to transfer mass.  In our analysis,
we discriminate between two types of collisions: those that result
from unstable mass transfer in a close binary system and those that
result from a dynamical interaction.  The latter case we identify as a
collision, whereas the former case is referred to as coalescence. In
this paper we focus on physical collisions between stars.

Our code employs ``standard N-body units''\footnote{For the definition
  of an $N$-body unit see {\tt
    http://en.wikipedia.org/wiki/Natural\_units\#N-body\_units}.}
(\cite{1986LNP...267..233H}), according to which the gravitational
constant, the total mass and the radius of the system are taken to be
unity.  The resulting unit of time is the $N$-body time-unit and is
related to the physical crossing time of the system through the
relation $T_{\rm cross} = 2 \sqrt{2}\,\tnb$.

\section{The circumstances of the first collision}\label{sect:Circum}
%
% We show data supporting the assumption that collisions 
% take place in the core of the cluster
%

In order to develop a better understanding of stellar collisions and
of the further evolution of the collision product, it is important to
know the conditions under which a collision takes place, such as
the mass, structure and composition of the participating stars, and the
geometry under which the collision occurs.

\subsection{The location of the first collision}
\label{sect:Numerical.Location}

Since the stellar density is highest in the cluster core, we expect
that majority of collisions to take place in the central region.  
\begin{figure}
  \begin{center}
    \includegraphics[scale=0.45]{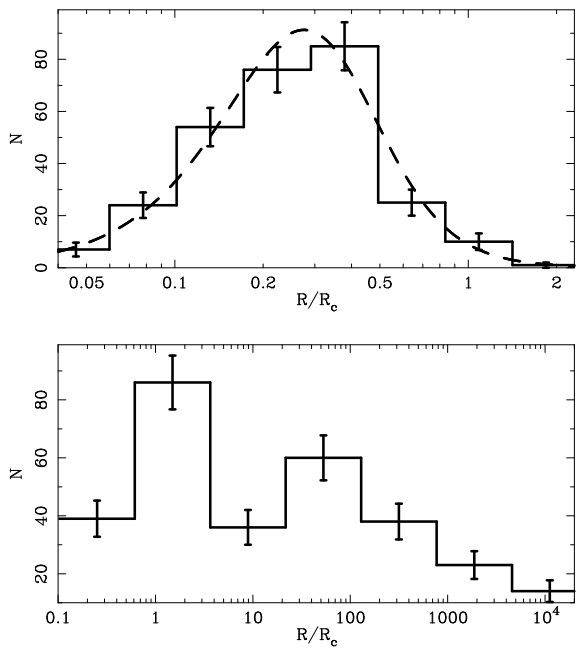}
  \end{center}
  \caption{Histogram of the number of collisions in all simulations as
    a function of the distance to the cluster centre in units of the
    instantaneous core radius. In the upper panel, the dashed line
    displays the fitted model of the number of collision. In the lower
    panel we present the distribution of coalescence as a function of
    distance to the centre of a cluster.}
  \label{fig:r_rcore_dist}
\end{figure}
In the top panel of \fig{\ref{fig:r_rcore_dist}}, we show the
distribution of the number of collisions as a function of the distance
to the cluster centre. All simulations presented in
\tbl{\ref{tbl:Initial_conditions}} are included in the sample.  Out of
a total of 282 collision in 509 simulations, only 33 occur outside the
instantaneous core of the cluster.

%% +++ v1 ++++
Though a small fraction ($12$\,\%), it is interesting that a sizeable
number of collisions occurs outside the core of a cluster. This can be
naturally explained as follows. In a cluster with a density profile
$\rho(r)$, the expected number of collisions in a spherical shell
located at radius $r$ is $N\propto r^2\rho(r)^2$, and therefore one
may expect a small, but finite, number of collisions to occur just
outside the core. To test this, we fitted this expression to the
number of collision. As a density profile, we used a variety of King
models with different values of $W_0$. We found that King models with
$W_0 \gtrsim 9$ fit well ($\chi^2 \approx 1$) since their density
profiles in the close neighbourhood of the core are essentially the
same.
%% +++

In the lower panel we present the distribution of number of
coalescence as a function of the distance to the cluster centre. We
see that the coalescence can occur quite far from the core of the
cluster; such coalescence are ejected binary stars in which the
massive companion leaves the main-sequence. The number of coalescence
far away from the core decreases as a function of distance to the
cluster centre.

\subsection{The time of the first collision}\label{sect:Numerical.Time}

\begin{figure}
  \begin{center}
    \includegraphics[scale=0.45]{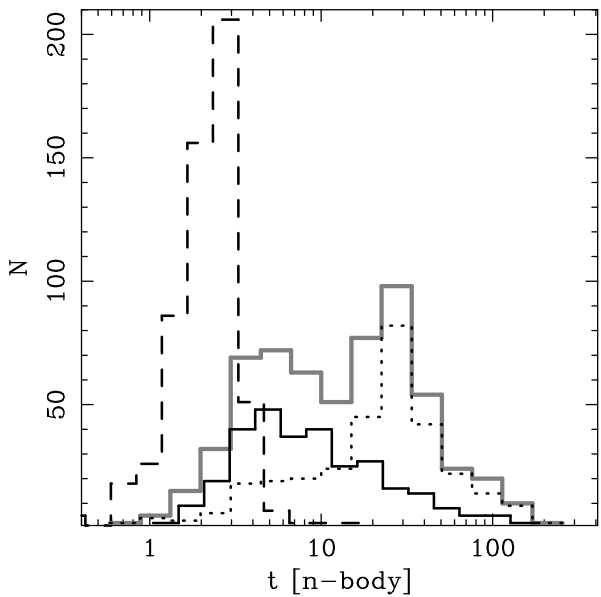}
  \end{center}
  \caption{Histogram of the time (expressed in $N$-body units) of
    formation of the first $|E| > 100$\,kT binary (dashed line), and
    of occurrence of the first collision (solid line) or first
    coalescence (dotted line). The total number of mergers (collisions
    + coalescence) in a time-bin is shown with a thick gray solid
    line.}
  \label{fig:tbin_tcoll_hist}
\end{figure}

In \fig{\ref{fig:tbin_tcoll_hist}}, we show the time distributions for
the formation of the first hard binary and for the first collision.
The first collision occurs preferentially after the formation of the
first hard binary, but the distribution is broad and extends all the
way to $\sim 200 N$-body time units. We notice that the coalescence
dominate at $t\gtrsim 20$ $N$-body units, whereas collisions are
dominant earlier.

The evolution of the mass function in the cluster centre is mainly
driven by dynamical friction, which preferentially brings massive
stars to the core. As a result, the mass function in the core becomes
flatter with time, and after core collapse the mass function stops
evolving except for the effects of stellar evolution, such as decrease
of the number of massive stars in a star cluster
\citep{2007MNRAS.378L..29P}.

\begin{figure}
  \begin{center}
    \includegraphics[scale=0.45]{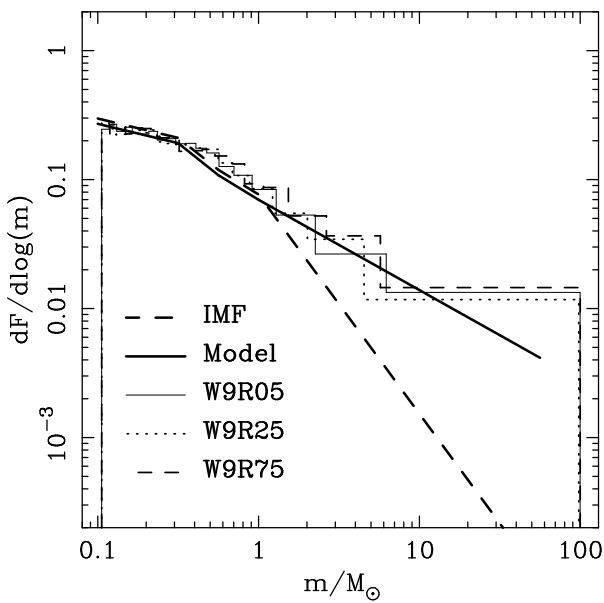}
  \end{center}
  \caption{Mass function in the core at the moment of the first
    collision averaged over all runs for each model. The thick dashed
    line represents the IMF whereas the thick solid line represents
    \eq{\ref{eq:core_mf}} which satisfactory represents the simulated
    mass function in the core.}
  \label{fig:core_mf}
\end{figure}

In \fig{\ref{fig:core_mf}} we present mass functions in the core at
the moment of the first collision averaged over all simulations in
which a collision happens. The mass functions for different models are
consistent with a single distribution better than at a 25\% level.
However, for models W9R50 and W9R75 the consistency is less than at
5\% level, which is due to the effects of stellar evolution.
% +++ v1 +++
As the ratio of a stellar evolution timescale to a dynamical timescale
is inversely proportional size of a star cluster, the effects of
stellar evolution become increasingly important as the size of the
cluster increases. While stellar evolution does not have a notable
influence on models W9R05, W9R10 and W9R25, it clearly leaves an
imprint in models W9R50 and W9R75 (see also
\append{\ref{sect:Appendix_A}}).
% +++

The mass function in the core after the first collision can be
approximated by the following expression:
\begin{equation}
  \mf_{\rm c}(m) \propto \left\{
  \begin{array}{ll}
    \mf_{\rm IMF}(m), & \textrm{if } m < m_1 = 2\mean{m_{\rm IMF}}, \\
    m\,\mf_{\rm IMF}(m), & \textrm{otherwise}.
  \end{array}
  \right.
  \label{eq:core_mf}
\end{equation}
Here $\mf_{\rm IMF}(m)$ is the initial mass function and $\mean{m_{\rm
IMF}}$ is the average mass of the initial mass function.  This
expression is presented as the thick solid line in
\fig{\ref{fig:core_mf}}.  Equation~\ref{eq:core_mf} implies that for
stars more massive than $\sim 2\mean{m_{\rm IMF}}$ the slope of the
mass function flattens. This is not unexpected since the dynamical
friction time-scale is inversely proportional to $m$. We conclude that
dynamical friction is a crucial ingredient in understanding the first
collision.

\begin{figure*}
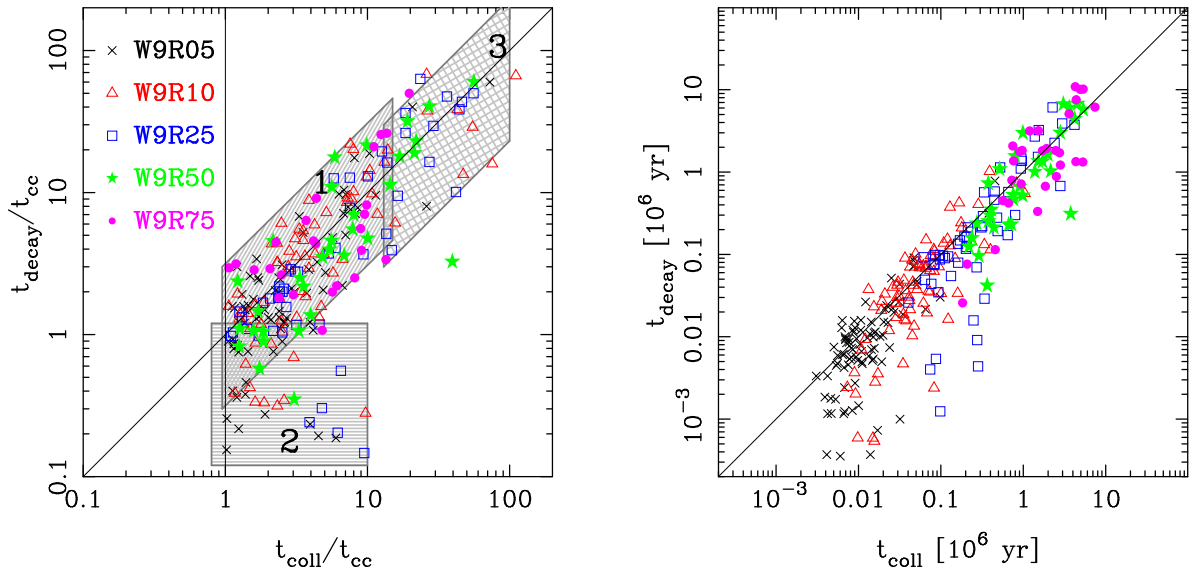

  \begin{center}
    \includegraphics[scale=0.45]{plots/t_decay_nbody.ps} $\quad\qquad$ 
    \includegraphics[scale=0.45]{plots/t_decay_phys.ps}
  \end{center}
  \caption{Correlation between \tcoll and \tdecay. In the left panel
    \tcoll and \tdecay are given in $N$-body units while in the right
    panel they are in physical units. The results from different
    models are indicated with different symbols and colours.  The
    diagonal in both panels gives the line for which \tcoll =
    \tdecay. The gray shaded areas in the left panel, numbered 1, 2
    and 3, indicate three different regimes, as described in the
    text. }
  \label{fig:t_decay_vs_t_coll_correlation}
\end{figure*}

In \fig{\ref{fig:t_decay_vs_t_coll_correlation}} we compare the time
of the first collision, \tcoll, with the time-scale on which a star
with mass $m$ sinks to the cluster centre from its initial orbit,
\tdecay.

We compute \tdecay\, for the most massive star that participates in
the first collision by integrating its equations of motion from the
initial orbit until the star decays to the cluster centre.  We include
the effect of dynamical friction using $\log(\Lambda) =
\mathrm{max}(0, \log(0.4M_{\rm cl}(r)/m))$, where $M_{\rm cl}(r)$ is
the cluster mass enclosed within a sphere of radius $r$
\citep{1987gady.book.....B}. As for the background potential, we adopt
a King $W_0=9$ density profile which represents our initial simulation
model.

The correlation between \tcoll, which is taken directly from the
simulations, and \tdecay, which is calculated as described above, is
presented in \fig{\ref{fig:t_decay_vs_t_coll_correlation}}.  In the
left panel the results are given in units of the core collapse time as
measured in the simulation under consideration, whereas in the right
panel time is given in physical units.

In the left panel of \fig{\ref{fig:t_decay_vs_t_coll_correlation}} we
identify three different regimes.  The majority of collisions are
distributed along $\tcoll = \tdecay$ (area 1 in
\fig{\ref{fig:t_decay_vs_t_coll_correlation}}).  This indicates, as we
suggested earlier, that dynamical friction is dominant in determining
the moment of the first collision.  A collision occurs quickly upon
the arrival of the star in the core.  

The dispersion along \tdecay\, in area 1 of
\fig{\ref{fig:t_decay_vs_t_coll_correlation}} is in part a consequence
of our assumption that the background potential is static throughout
our calculations of \tdecay.  The density profile of the clusters,
however, is calculated self consistently in our $N$-body simulations,
and it changes with time.  In addition, even when the star arrives in
the core, it still takes some time before it engages in a collision.
The latter effect is visible in area 2 of
\fig{\ref{fig:t_decay_vs_t_coll_correlation}}.

In area 2 of \fig{\ref{fig:t_decay_vs_t_coll_correlation}}, we present
the stars that experience a collision later than one would naively
expect from their calculated decay time.  These stars are born in or
near the cluster centre but a collision is delayed up to the moment of
core collapse. After that, the stars still have to participate in a
strong encounter which leads to a collision. Thus, the moment of the
collision is determined by $\tcc$ and the time required to find a
suitable collision candidate, and this may take up to about 10\,\tcc.

It is somewhat surprising that there is a third region along the line
$\tcoll = \tdecay$, which is illustrated in the area 3 of
\fig{\ref{fig:t_decay_vs_t_coll_correlation}}, and which extends all
the way to $\tcoll \simeq 100\,\tcc$. As a rule of thumb, it takes
roughly 50 $N$-body units for a 50\msun\, star to decay from the
half-mass radius to the core. From area 2, it can be seen that a star
in the core may require up to 10\,\tcc, which roughly corresponds to
30 $N$-body time units (see \fig{\ref{fig:tbin_tcoll_hist}}), to
engage in a collision. Since we expect our simulations to host at
least one $50 \msun$\, star which is initially located within the
half-mass radius, it is rather unlikely to have collisions in our
simulations after $t \simeq 15\tcc$; still, some collisions do happen
as late as 100\,\tcc.

These late collisions, which are illustrated in area 3 of
\fig{\ref{fig:t_decay_vs_t_coll_correlation}}, are attributed to
massive stars that reach the core but instead of experiencing a
collision are ejected from the cluster in a strong encounter with a
binary. This effectively delays the moment of the first collision
since the potential target star is removed from the cluster. If this
happens, the first collision is postponed until the moment when
another massive star reaches the cluster core and subsequently
collides.  In \append{\ref{sect:Appendix_B}}, we show that such
self-ejections are possible in clusters with mass
\begin{equation}
  M_{\rm cl} \gtrsim 2\cdot 10^4
  \left(\frac{m_\star}{50\,\msun}\right)^3
  \left(\frac{m_{\rm s}}{10\msun}\right)^{-2}\,\msun.
  \label{eq:m_se}
\end{equation}
Here, $m_{\star}$ is the mass of an ejected star, and $m_{\rm s}$ is
the mass of the star triggering the ejection event.

\begin{figure*}
  \begin{center}
    \includegraphics[scale=0.8]{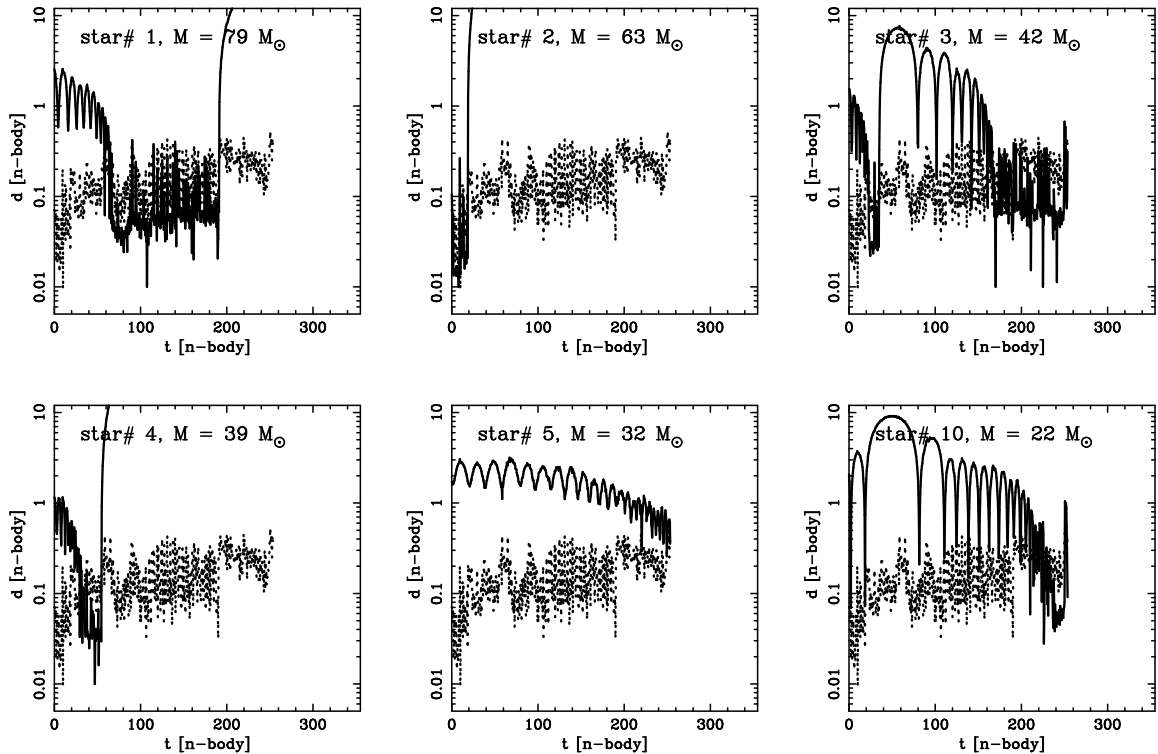}
  \end{center}
  \caption[]{Distance to the cluster centre for six of the most
  massive stars in one of the simulations from model W9R05.  The
  dotted curve in each panel gives the evolution of the core radius of
  the model (which is identical in each panel).  The solid curve gives
  the evolution of the distance to the cluster centre for the star
  identified in the top left corner of the panel.}
  \label{fig:W9_run593_stars} 
\end{figure*}
The process described in the previous paragraph
%
%relation to area 3 in \fig{\ref{fig:t_decay_vs_t_coll_correlation}} 
%
is illustrated in \fig{\ref{fig:W9_run593_stars}} where we present the
evolution of the distance to the cluster centre for several of the
most massive stars in one of the simulations of W9R05 (see
\tbl{\ref{tbl:Initial_conditions}}).  In addition, we plot the
evolution of the core radius in each panel.

The most massive cluster member, star \#1 of $79${\msun}, which is
presented in the top left panel of \fig{\ref{fig:W9_run593_stars}},
sinks from about the half-mass radius to the cluster core in roughly
50\,$N$-body time units ($\simeq 15$\tcc in this particular run). It
becomes a binary member at $t\simeq 88.5$ and the binary increases its
binding energy to $\sim$ 100\,kT at $t\simeq 165$ $N$-body time
units. Star \#1 stays in the core until it is ejected at about
200\,$N$-body time units, never to return again. Even though star \#1
is the most massive star in the system, it is part of a binary system
and it resides in the cluster centre for more than $100 N$-body time
units, it does not participate in a collision but is ejected from the
core.

The same process causes several of the other massive stars to be
ejected, such as stars \#2 and \#4 in \fig{\ref{fig:W9_run593_stars}},
whereas some of the other massive stars, such as stars \#3 and \#5,
are not ejected.  These repeated ejections of high-mass stars delay
the collision until nearly 250\,$N$-body time units, which roughly
corresponds to 80 \tcc.  Eventually, it is 22\msun star \#10 which
reaches the core and experiences a collision with a 4\msun\, star.

\subsection{Mass distributions}\label{sect:Numerical.Masses}

The binary which forms during core collapse is likely to be the
candidate for a collision. However, this does not mean that the two
stars in the binary coalesce, instead this enhances the probability
for a collision with a third star.

In \tbl{\ref{tbl:merge_outcome}} we present the number of collisions
that occurred in each of the models with respect to the choreography
of the triple interaction. The notation is as follows: the two binary
members are called $M_p$ and $M_s$ for the most massive (primary) and
least massive (secondary) star respectively, while the third star is
called the bullet and is indicated with $M_b$; the two colliding stars
are presented in braces while the entire triple interaction is in
parenthesis.

\begin{table*}
  \begin{tabular}{cccccc}
    \hline 
    Model & $(\{M_p,M_b\},M_c)$ & $(\{M_s,M_b\},M_p)$ & $(\{M_p,M_s\},M_b)$ & $N_{\rm c}$ & $N_{\rm m}$ \\
    \hline
    W9R05  &  67     & 15   & 5    & 87 & 13 \\
    W9R10  &  59     & 11   & 9    & 79 & 21 \\
    W9R25  &  37     & 12   & 5    & 54 & 46 \\
    W9R50  &  16     & 12   & 5    & 33 & 66 \\
    W9R75  &  19     & 5    & 5    & 29 & 81 \\
    \hline
  \end{tabular}
  \caption[] {The choreography of triple interactions leading to a
    collision in the different calculations.  In the first column, we
    present the model name followed by the number of collisions in
    each of the configurations.  These are: a collision between the
    primary and the bullet (column two), a collision between the
    secondary and the bullet (column three) and a collision between
    the primary and the secondary star (column four).  The fifth
    column shows the total number of collisions that are outcomes of a
    dynamical interaction, whereas the last column shows the number of
    binary mergers which result from an unstable phase of mass
    transfer in a dynamically formed binary. The latter category is
    not further discussed in this paper.}
  \label{tbl:merge_outcome}
\end{table*}

The collisions in the densest star clusters (model W9R05 and W9R10)
are dominated by collisions between the primary and the bullet star.
The fraction of these collisions remains roughly constant compared to
the total number of collisions. The shallowest clusters (model W9R50
and W9R75), on the other hand, are governed by binary evolution, which
does not come as a surprise since stellar evolution plays an
increasingly important role as the size of the cluster increases. In
some cases, however, the primary is not participating in the
collision, but instead it is the secondary star that collides with the
bullet.  

% +++ 

\begin{figure*}
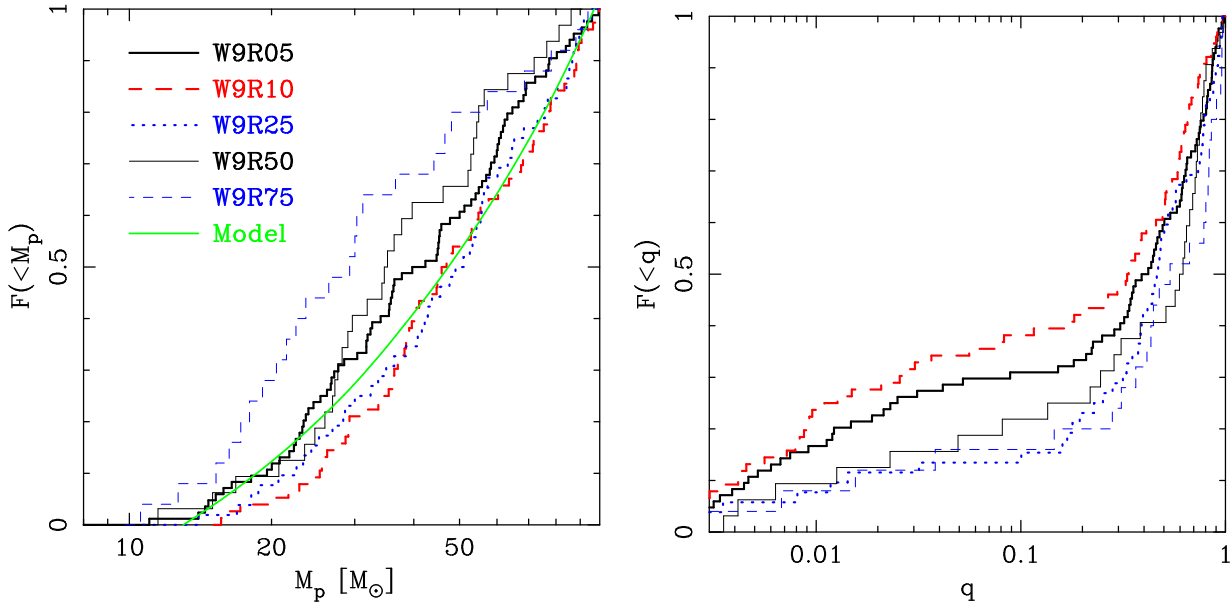

  \begin{center}
    \includegraphics[scale=0.45]{plots/mf_coll_prim.ps}$\quad$
    \includegraphics[scale=0.45]{plots/q_coll.ps}
  \end{center}
  \caption{ Distribution of primary masses (left) and mass ratios
    (right) for binaries participating in the collision. According to
    Kolmogorov-Smirnov test there is 15\%, 28\%, 90\%, 5\% and 0.2\%
    chance for models W9R05, W9R10, W9R25, W9R50 and W9R75
    respectively that the deviations in the distributions from the
    theoretical curve are random in nature. The green solid line is a
    cumulative mass function in the core
    (\append{\ref{sect:Appendix_A}}), but with the lower mass limit
    taken to be $15\msun$ .}
  \label{fig:primary_mass_ratio_coll}
\end{figure*}

In the left panel of \fig{\ref{fig:primary_mass_ratio_coll}}, we
present the distribution of primary masses. These distributions are
statistically indistinguishable for the three densest clusters (models
W9R05, W9R10 and W9R25), whereas for the shallowest clusters (models
W9R50 and W9R75) they deviate in that the mean mass for the primary
stars decreases. This is the result of stellar evolution, which
becomes gradually more important for shallower clusters. The mass
functions of the primary in models W9R05, W9R10 and W9R25 are
consistent with the mass function in the core at the moment of the
first collision (\append{\ref{sect:Appendix_A}}), if only stars above
$15\,\msun$ are taken into account.

\begin{figure}
  \begin{center}
    \includegraphics[scale=0.45]{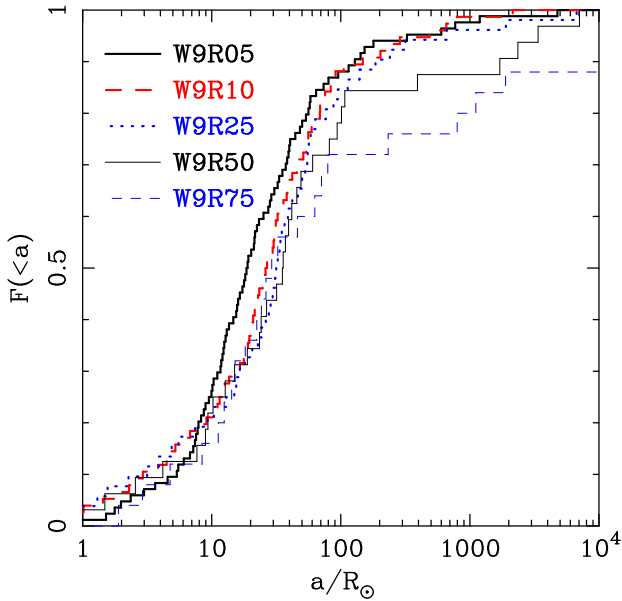}
  \end{center}
  \caption{ Distribution of semi-major axes for binaries undergoing a
    collision. Binaries with separation smaller than the sum of their
    stellar radii $a \lesssim 10\,\rsun$ are contact binaries which
    form in the course of the simulation.}
  \label{fig:a_dist}
\end{figure}

In the right panel of \fig{\ref{fig:primary_mass_ratio_coll}}, we
present the distribution of the mass ratio of the secondary to the
primary star. Here, we see that the mass ratio for shallower clusters
is systematically higher than for denser clusters. This trend we
explain by the fact that the binaries in shallower clusters experience
more interactions before they participate in a collisions, allowing
for exchanges of a more massive star into the binary. This is
supported by \fig{\ref{fig:a_dist}} where we plot the distribution of
orbital separations of the binaries that participate in a collision
event. These distributions are statistically indistinguishable when
displayed in physical units. This however implies that the binaries in
the larger clusters are harder as their semi-major axis is smaller
when measured in $N$-body units.

\begin{figure}
  \begin{center}
    \includegraphics[scale=0.45]{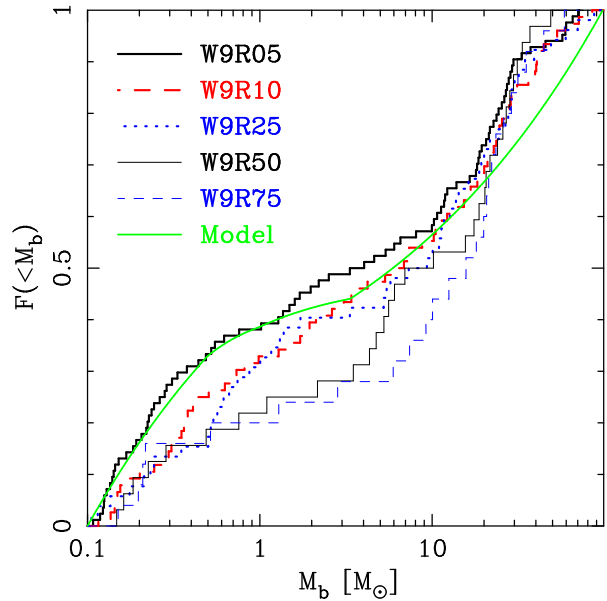}
  \end{center}
  \caption[]{ Mass function of single stars that collide with
    binaries. The solid line shows a cumulative distribution function
    computed from a bullet mass function presented in
    \eq{\ref{eq:bullet_mf}}}
  \label{fig:bullet_mf}
\end{figure}

In \fig{\ref{fig:bullet_mf}} we present the distribution of the masses
of bullet stars colliding with the primary. The mean mass of bullet
stars increases with the size of the cluster. Together with the
simulation data, we present the theoretical line which gives the
results of a qualitative model for the mass of the bullet star. The
low-mass end of this curve follows the mass function in the cluster
core at the moment of the collision, whereas for the steeper high-mass
part (above $\sim 2\langle m \rangle_{\rm core}$), we weight the
probability distribution with the gravitational focusing of the
bullet:
\begin{equation}
  \mf_{\rm b}(m_{\rm b}) \propto \left\{
  \begin{array}{ll}
    \mf_{\rm c}(m_{\rm b}), & \textrm{if } m_{\rm b} < 2\mean{m}_{\rm core}, \\
    m_{\rm b}\,\mf_{\rm c}(m_{\rm b}), & \textrm{otherwise}.
  \end{array}
  \right.
  \label{eq:bullet_mf}
\end{equation}
We make the distinction between the enhanced cross section (high-mass
end) and the geometric cross section (low-mass tail), since we expect
that for low-mass bullet stars the collision rate is dominated by the
cross section of the encountering binary, rather than the bullet
itself.
% \simon{Evghenii, please check, think of a way to check this.}

\section{The collision geometry}\label{sect:CollisionGeometry}
In the previous sections, we demonstrated that all collisions in a
cluster's centre occur between a binary component and a single
star. This has far reaching consequences for the energetics and
angular momentum of the collision. While in a two-body collision the
outcome of the event depends only on the impact parameter and the
relative velocity at large separation, in our simulations the
situation is considerably more complicated as one of the encountering
objects is always a binary member. In this case the relative velocity
at the moment of the impact can be either significantly higher or
lower than in the idealised two-body case. Therefore, the consequences
for the evolution of the collision product may be profound.

\begin{figure*}
  \begin{center}
    \includegraphics[scale=0.45]{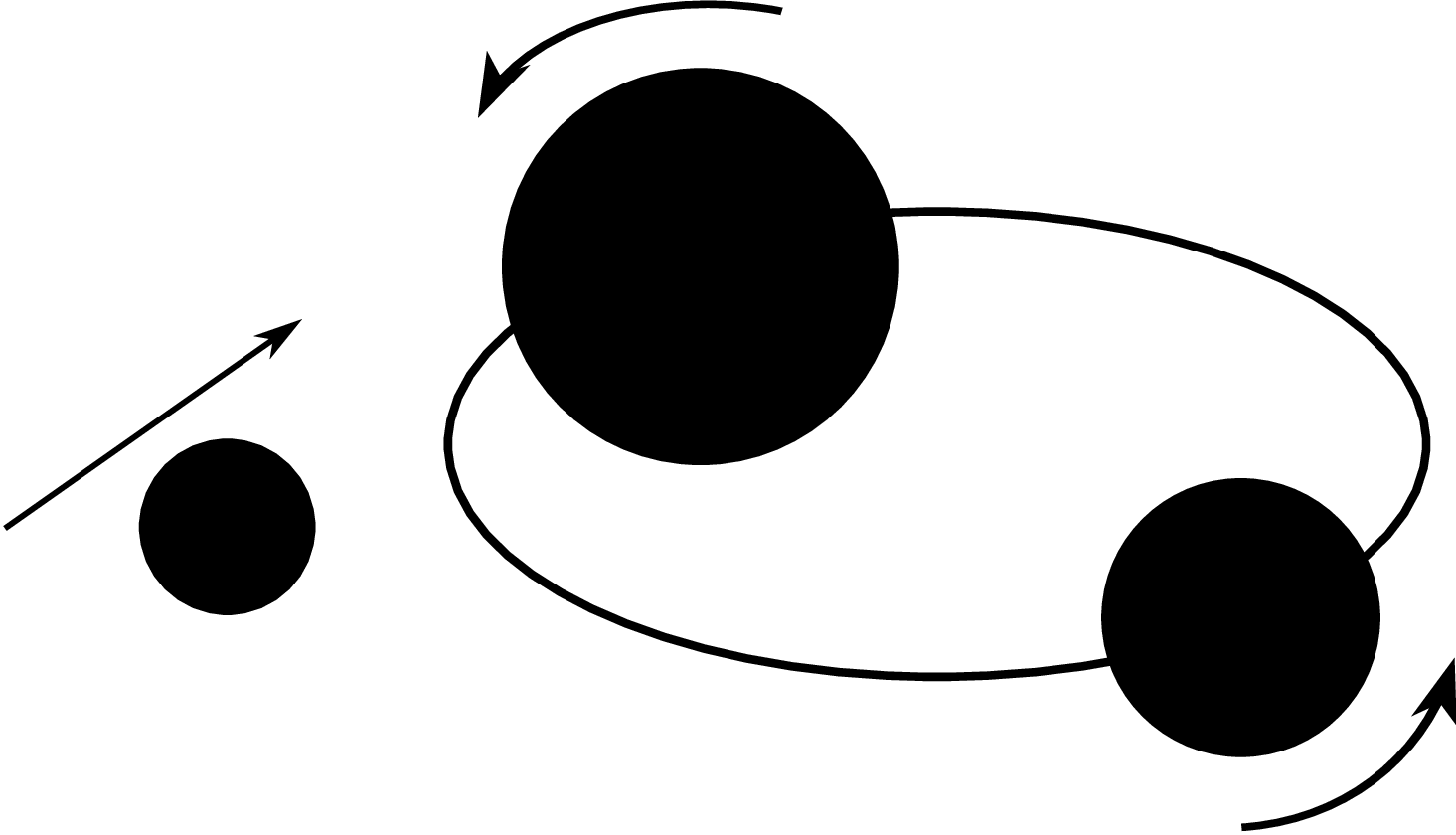}
    $\qquad$
    $\qquad$
    $\qquad$
    \includegraphics[scale=0.45]{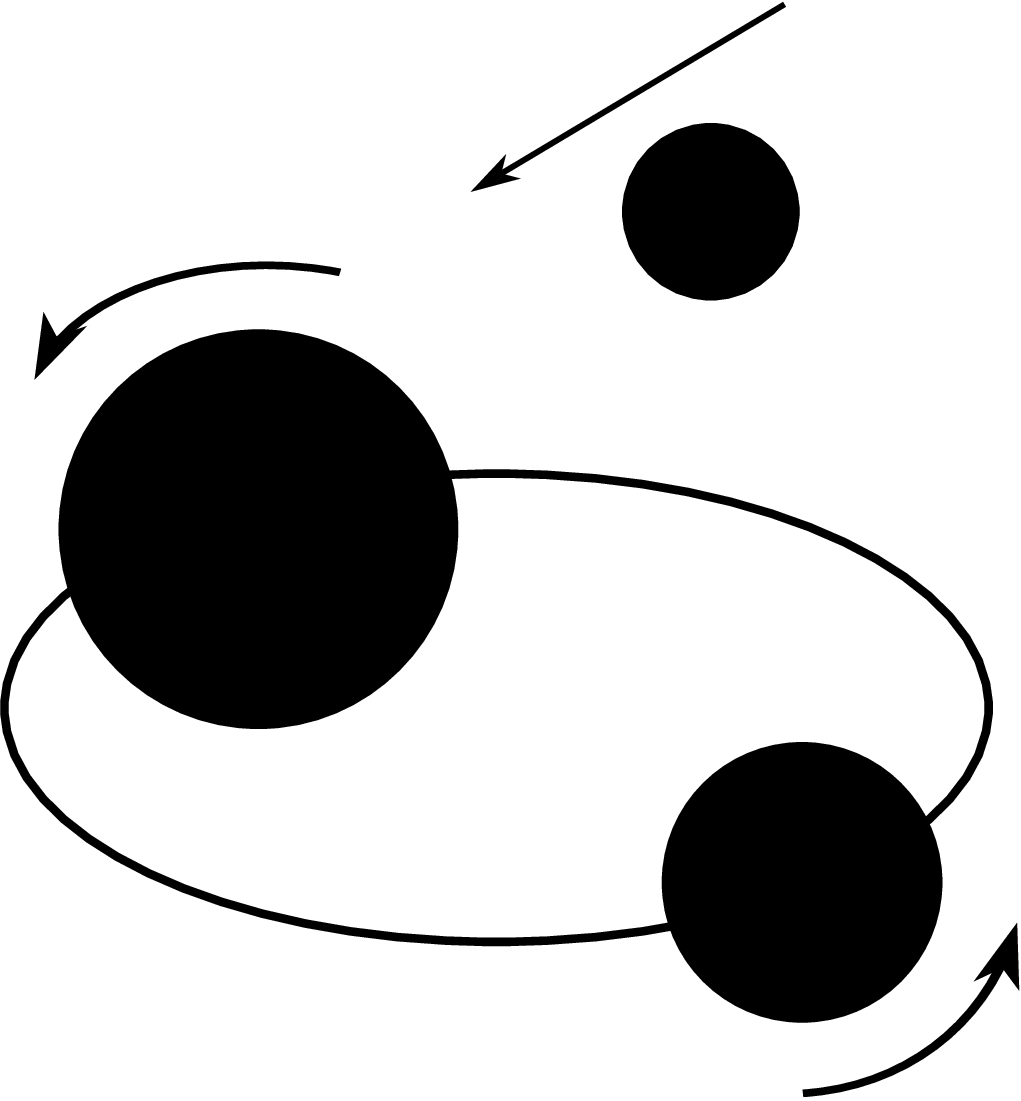}
  \end{center}
  \caption[]{Schematic representation of the collision between a
    binary member and a single star. The left figure shows the case
    when the two colliding stars move towards each other. In this
    instance, the impact velocity can exceed the escape velocity from
    the two body systems formed by the colliding stars. The right
    figure shows the case when the two interacting stars move in the
    same direction. In this case, if a collision occurs, the impact
    velocity can be smaller than the escape velocity.}
  \label{fig:imp_vel_s}
\end{figure*}

\fig{\ref{fig:imp_vel_s}} illustrates the two extreme cases that can
occur when a bullet star collides with a star in a binary.  In the
left panel we show two colliding stars that are approaching each other
at the moment of the first contact, whereas in the right panel we show
two colliding stars that are moving in the same direction at the
moment of impact, so that the binary companion is effectively receding
from the bullet star.

\begin{figure}
  \begin{center}
    \includegraphics[scale=0.45]{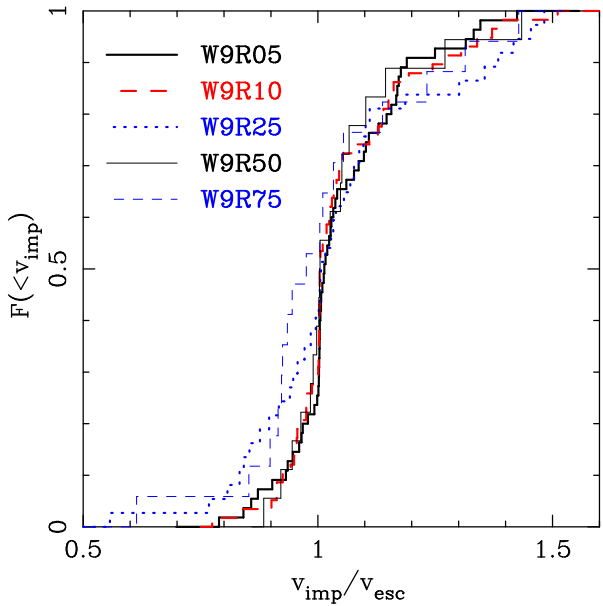}
  \end{center}
  \caption[]{Cumulative distribution of impact velocities in units of
    the escape velocity of the two-body system formed by the colliding
    stars. The thick solid line corresponds to a W9R05 model, the
    thick dashed line to a W9R10, the thick dotted line to a
    W9R25, the thin solid line to a W9R50 and the thin dotted line
    to a W9R75 model.}
  \label{fig:v_infty_hyper}
\end{figure}

The consequence for the impact velocity in terms of the escape speed
of the merged object is illustrated in
\fig{\ref{fig:v_infty_hyper}}. 
Here, we see that about half of the collisions occur with a velocity
smaller than the one expected for a two-body encounter. In some
extreme cases, however, the velocity at impact can be $\simeq 50$\%
higher than in the two-body case. The tail of lower impact velocities
is completely absent in isolated two-body unbound encounters; this
latter case is somewhat comparable to the merger of two binary
components.

\begin{figure}
  \begin{center}
    \includegraphics[scale=0.45]{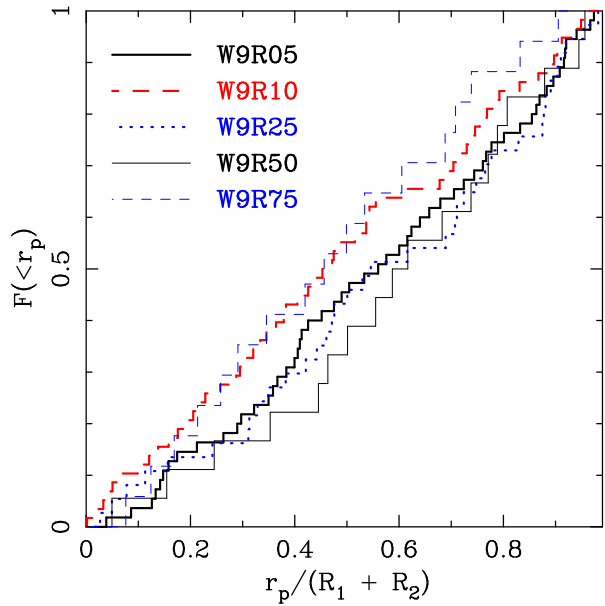}
  \end{center}
  \caption[]{ Cumulative distribution of pericentres. The periscentres
    are computed assuming that two stars approach each other on a
    hyperbolic trajectory such that their relative velocity at contact
    is equal to the one observed in the simulations.}
  \label{fig:r_p_hyper}
\end{figure}

In \fig{\ref{fig:r_p_hyper}}, we present the distance between the two
stellar centres at the moment of impact. 
For all models, the distribution of impact distance is flat and as a
result the cumulative distribution is a straight diagonal line. This
is a direct consequence of gravitational focusing, which dominates
even in these three-body encounters.  The chance to have a relatively
small impact distance is comparable to the chance of having a very
large impact distance, which is against our naive intuition that the
probability of a collision $\propto R_\star^2$.

\section{Discussion and conclusions}
\label{sect:Discussion}

We have studied the dynamics of the first stellar collision occurring
in the evolution of young star clusters by means of high-accuracy
direct $N$-body simulations. We have carried out about 500
simulations of star clusters represented by King models
of different central concentration and different size.

During the early evolution of young dense clusters, massive stars sink
to the core due to dynamical friction. As a result, the core becomes
enhanced with massive stars, which can be seen in the flattening of
the core mass function for $m \gtrsim 2\mean{m}$.  Due to the steeper
dependence of the binary formation cross-section on stellar mass than
the collision cross-section, binary formation becomes a more likely
process than a collision.  Young clusters are, in this respect,
different from globular clusters, where binary formation by three-body
encounters is unimportant \citep{1983Natur.301..587H}.

Collisions occur after the formation of hard binaries in the core of
the cluster. As expected, nearly all collisions occur in the core of
the star cluster.  The time of the first collision is roughly equal to
the time required for the most massive colliding star to reach the
core. While most of the collisions occur within a timescale of 50
$N$-body time units (15\tcc), we find that some of them are delayed to
as much as a few hundred $N$-body time units. This is due to the fact
that some of the massive stars that have reached the core are ejected
from the cluster during dynamical encounters. This process postpones
the first collision event to later times.

As for the geometry of the collision, we have found that most of the
collisions occur between the primary star of a participating binary
and a single star. The fraction of these collisions remains constant
as the size of the cluster changes, except for the models W9R50 and
W9R75 which are affected by stellar evolution. The masses of the
primary star are distributed according to the core mass function but
only for stars with $M_p \gtrsim 10\,\msun$. The bullet stars, on the
other hand, are single core stars.

One of the consequences of this geometry is that the impact velocity
covers a wide range of values: from roughly 50\% to 150\% of the
escape velocity from the two-body system formed by the two colliding
stars. On the one hand, the low velocity tail of the impact velocity
would be impossible if collisions were to occur between two unbound
stars; however, this situation is to some degree similar to the merger
of a binary. On the other hand, the high velocity tail can only occur
if the dispersion velocity in the system is comparable to the escape
velocity from the stellar surface. Such high velocity collisions may
result in a significant mass loss or even destruction of a star
\citep{2005MNRAS.358.1133F}. As for the impact parameter, we have
found that all collisions are dominated by gravitational focusing. The
distribution of pericentre separations implies that nearly head-on
collisions are as frequent as off-axis collisions.

As the collisions occur in a binary-single stellar system, there is a
possibility that all there stars may merge. If a single star merges
with one of the binary members, the resulted collision product can
expand by a large factor due to excess of thermal energy. In this
case, the binary may become unstable and merge, and this therefore
results in a triple merger. On the other hand, if the first encounter
which involves a single star and a binary companion is not a head-on
but rather grazing one, the possibly large impact velocity may prevent
the merger all together. Therefore, to understand the fate of such
systems one has to resort to hydrodynamic simulation.

\section*{Acknowledgements}

This work was supported by NWO (grants \#635.000.303 and
\#643.200.503), NOVA, the LKBF. AG is supported by grant NNX07AH15G
from NASA. The calculations for this work were done on the MoDeStA
computer in Amsterdam, which is hosted by the SARA supercomputer
centre.

\bibliographystyle{mn2e}
\bibliography{GGPZ2006}

\appendix

\section{Mass function of binaries}\label{sect:Appendix_A}

In a given stellar population, it is possible to estimate the mass
function of binary stars formed by three-body encounters. Let $\mf(m)$
be the mass function of single stars in the region where binary
formation takes place. Our aim here is to estimate the mass function
of binaries formed by three-body encounters, $\mf_{bin}(M_p,q)$. We
wish to express this as a function of the mass of the primary star,
$M_p$, which is the most massive binary companion, and the mass ratio,
$q < 1$, of the secondary, which is the least massive binary
companion, to the primary.

The probability to form a binary with a star of mass $m_1$ and another
star of mass $m_2$ is proportional to the product of probabilities to
randomly draw these stars from a mass function, $\mf(m_1)\mf(m_2)$,
and the cross-section for these two stars to form a binary,
$\Sigma(m_1, m_2)$. However, to form a binary a third star is required
which carries away energy in order for $m_1$ and $m_2$ to form a bound
system. In the further analysis, we assume that the mass of the third
star is small compared to $m_1$ or $m_2$ and, therefore, it can be
neglected in the cross-section of binary formation by three-body
encounters.
 
Following \cite{2003gmbp.book.....H}, we write the binary formation
cross section in the following form
\begin{equation}
  \Sigma(m_{1},m_{2})_b\propto (m_{1} + m_{2})/v_{12}^2.
  \label{eq:sigma_binary_formatin_general}
\end{equation}
Here, $v_{12}$ is the relative velocity between two stars. Assuming
energy equipartition, we write $v_{12}^2 \propto (m_1+m_2)/(m_1m_2)$,
and \eq{\ref{eq:sigma_binary_formatin_general}} takes the following
form
\begin{equation}
  \Sigma_{b}(m_{1},m_{2})\propto m_1m_2.
  \label{eq:sigma_binary_eq}
\end{equation}
As we have mentioned above, the mass function of the dynamically
formed binaries is
\begin{equation}
  dF\propto dm_{1}dm_{2}{\cal N}(m_{1}){\cal
    N}(m_{2})\Sigma_b(m_{1},m_{2}).
\end{equation}
After the change of variables from $(m_1, m_2)$ to $(M_p, q =
M_s/M_p)$, were $M_p={\rm max}(m_1,m_2)$ and $M_s ={\rm
  min}(m_1,m_2)$, this equation takes the following form
\begin{equation} 
  \frac{dF}{dqdM_p}={\cal N}_{bin}(M_p,q)\propto qM_p^{3}
       {\cal N}(M_p) \mf(qM_p).
  \label{eq:binary_df}
\end{equation}

It is now possible to find the distribution of $M_p$ and of $q$. The
former can be obtained by integrating \eq{\ref{eq:binary_df}} over
all possible mass ratios
\begin{equation}
  \mf_{bin}(M_p) = \int dq\,\mf_{bin}(M_p,q),
  \label{eq:Nbin_Mp}
\end{equation}
and the latter is the integral of \eq{\ref{eq:binary_df}} over all
primary masses
\begin{equation}
  \mf_{bin}(q) = \int dM_p\,\mf_{bin}(M_p,q).
  \label{eq:Nbin_q}
\end{equation}

We compare both \eq{\ref{eq:Nbin_Mp}} and \eq{\ref{eq:Nbin_q}} with
the simulations. Given the fact that most of the collisions occur in
the core of a star cluster (see \sect{\ref{sect:Numerical.Time}}), we
assume that binaries also form in the core. It was shown by
\cite{2007MNRAS.378L..29P} that the mass function in the core is unchanged after
the formation of the first hard binary. Thus, we assume that
\eq{\ref{eq:core_mf}}, which is the mass function at the moment of
collision, is the mass function in the core of the star cluster at the
moment of the formation of the first hard binary, $\mf(m)$.

We extract the mass function of binaries which are formed by
three-body encounters in the following way. For each simulation, we
obtain the masses of the primary and the secondary star. Since some
binaries persist for a long period of time, we make sure that the same
binary is not used more than once. After that, we build distributions
for the masses of the primary star and the mass ratio.

\begin{figure*}
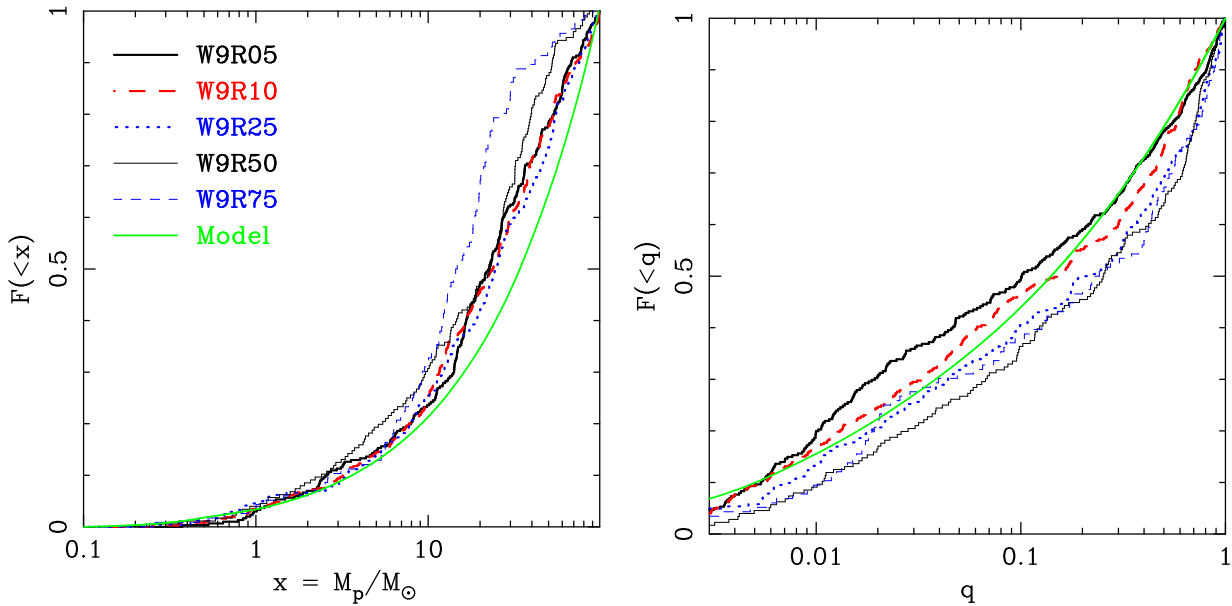

  \begin{center}
    \includegraphics[scale=0.45]{plots/bin_mp.ps}$\quad$
    \includegraphics[scale=0.45]{plots/bin_q.ps}
  \end{center}
  \caption{ Primary mass (left) and mass ratio (right) distribution
    for dynamically formed binaries. The mass distributions of primary
    stars for W9R05, W9R10 and W9R25 models are consistent with a
    single distribution function better than at 30\% level. The source
    of the discrepancy in the high-mass end of the mass function is
    due to the effects of stellar evolution on the initial mass
    function.  }
  \label{fig:primary_mass_ratio_binaries}
\end{figure*}

We show the resulting distributions, both from the simulations and from
our semi-analytical estimates, in
\fig{\ref{fig:primary_mass_ratio_binaries}}.
It may be noted that models W9R50 and W9R75 lack massive stars. This
is the result of stellar evolution that modifies the high-mass tail of
the initial mass function. The effect is less pronounced in the W9R50
model and is unnoticeable in the rest of the models.

\section{Binary self-ejection}\label{sect:Appendix_B}
In this appendix, we estimate the minimal mass of a star cluster that
can yield self-ejecting binaries.

Given a star cluster of mass $M$ and half-mass radius $R$, its
gravitational binding energy is
\begin{equation}
  E \simeq \frac{GM^2}{4R}.
  \label{eq:cluster_binding}
\end{equation}
For simplicity, in this analysis we consider a binary which consists of
two equal-mass stars of mass $m_\star$. The binding energy of such a
binary of semi-major axis $a$ is
\begin{equation}
  E_b \simeq \frac{Gm_\star^2}{2a}.
  \label{eq:binary_binding}
\end{equation}
In order to prevent a star cluster from collapse, the binary should be
able to generate enough heat. In this case, we assume that the binding
energy of the binary should be equal to the binding energy of the
cluster. Combining \eq{\ref{eq:cluster_binding}} and
\eq{\ref{eq:binary_binding}}, we show that this occurs when the
semi-major axis of the binary is
\begin{equation}
  a_{\rm eq} \simeq 2 R \left(\ \frac{m_\star}{M}\right)^2.
  \label{eq:a_eq}
\end{equation}

If a single stars with mass $M_{\rm s}$ star were to interact with
such a binary, it would be ejected with a velocity of the order of the
orbital velocity of the members of the binary, $v^2_{\rm orb} \simeq
Gm_\star/a_{\rm eq}$. In order to conserve linear momentum, the binary
itself would recoil with velocity
\begin{equation}
  v_{\rm rec} = \frac{m_{\rm s}}{2m_\star}v_{\rm orb}.
  \label{eq:vrec_binary}
\end{equation}
Here, we assume that the mass of the low-mass star is equal to the
mean stellar mass in the core, $\mean{m_c}$.

We estimate the escape velocity from the cluster in the following way
\begin{equation}
  v^2_{\rm esc} \simeq \frac{2GM}{R}.
  \label{eq:vesc_cluster}
\end{equation}
Combining \eq{\ref{eq:vrec_binary}} and \eq{\ref{eq:vesc_cluster}}, we
express the condition $v_{\rm rec} > v_{\rm esc}$ as a condition on
the minimal mass of the cluster which can yield ejected binary stars
\begin{equation}
  M \gtrsim 16\frac{m_\star^3}{m_{\rm s}^2} \approx
  2\cdot 10^4 \left(\frac{m_\star}{50\,\msun}\right)^3
  \left(\frac{m_{\rm s}}{10\,\msun}\right)^{-2} \,\msun.
  \label{eq:m_selfejection}
\end{equation}

\end{document}